# A Formulation of the Simple Theory of Types (for Isabelle)

*Lawrence C. Paulson*

Computer Laboratory, University of Cambridge

Pembroke Street, Cambridge CB2 3QG, England

2 August 1989

**Abstract**

Simple type theory is formulated for use with the generic theorem prover Isabelle. This requires explicit type inference rules. There are function, product, and subset types, which may be empty. Descriptions (the $\eta$-operator) introduce the Axiom of Choice. Higher-order logic is obtained through reflection between formulae and terms of type *bool*. Recursive types and functions can be formally constructed.

Isabelle proof procedures are described. The logic appears suitable for general mathematics as well as computational problems.

# Contents

# 1   Introduction

*Isabelle* is a theorem prover for various logics, including several first-order logics, Martin-Löf's Type Theory, and Zermelo-Fraenkel set theory [30, 31]. Each new logic is formalized within Isabelle's meta-logic. New types and constants express the syntax of the logic, while new axioms express its inference rules.

The present formulation of simple type theory (also called higher-order logic) may interest logicians. It also illustrates Isabelle applied to an area where hard choices must be made. The simple theory of types is *too* simple; it can be enriched in numerous ways, as hinted by Church [5]. The traditional formalization, with implicit type constraints, goes beyond Isabelle's view of syntax. Isabelle supports ML-style type inference with unification; the formulation offers a limited degree of polymorphism.

There are several reasons for implementing higher-order logic in Isabelle.

Gordon and others have used higher-order logic, with great success, for hardware verification [6, 14]. They have developed a theorem prover called HOL, based on LCF. Another implementation of higher-order logic is TPS [2]. How well does Isabelle perform against specialized systems like these?

Zermelo-Fraenkel set theory [36] is intended as a foundation of mathematics, but is inconvenient for formal proof — even set theorists use intuition and diagrams. Yet set theory is the basis of the Z specification language [35]. Philippe Noël has conducted extensive set theory proofs in Isabelle [25]. Type theory is also intended as a foundation for mathematics, and seems nearly as powerful. How does it compare with set theory as a practical formal language?

Isabelle's meta-logic is a fragment of Church's version of type theory [30]. It is natural to ask whether the meta-logic can be formalized in itself. Actually, the object version of higher-order logic has to be much larger than the meta-logic because it is intended for expressing all kinds of mathematics. The meta-logic only has to express other formal systems.

The Isabelle implementation is definitely not intended for teaching Church's notation. Here TPS is the champion, with a special character set for Church's subscripted Greek letters. However, most modern authors write $x : \alpha$ and $\alpha \to \beta$ rather than $x_\alpha$ and $\beta\alpha$. Church's axiom system is now antiquated, largely dating back to *Principia Mathematica*. There are improved formulations but most use the Hilbert style. Natural deduction is far superior for automated proof.

Sections 2 and 3 of the paper introduce Isabelle and the theory of types. Sections 4 to 10 discusses some issues in the Isabelle formulation, while Section 11 presents the formulation itself. The remaining sections describe proof procedures and offer some conclusions. The computer file containing the rules is included as an appendix.

## 2  Overview of Isabelle

Isabelle represents its object-logics within a fragment of intuitionistic higher-order logic including implication, universal quantifiers, and equality [30]. The implication $\phi \implies \psi$ means '$\phi$ implies $\psi$', and expresses logical entailment. The quantification $\bigwedge x \,.\, \phi$ means '$\phi$ is true for all $x$', where $x$ has a fixed type, and expresses generality in object-rules and axiom schemes. The equality $a \equiv b$ means '$a$ equals $b$', and expresses definitions.

The meta-logic includes the typed $\lambda$-calculus, which is convenient for formalizing the syntax of object-logics, particularly variable binding. Provisos of quantifier rules (of the sort '$x$ not free in the assumptions') are enforced by meta-level quantification.

Like in LCF [29], backwards proofs are developed using tactics and tacticals, which are implemented using Standard ML. But an inference rule in LCF is a function from the premises to the conclusion, while in Isabelle it is an axiom in the meta-logic stating that the premises imply the conclusion.

Since Isabelle axioms are essentially Horn clauses, the proof techniques draw ideas from PROLOG. Huet's higher-order unification procedure [18] takes account of $\alpha$, $\beta$, and $\eta$-conversions during unification. Higher-order unification can return multiple or infinitely many results. While the general problem is undecidable, the procedure works well in Isabelle.

## 3  A brief history of type theory

Bertrand Russell invented the theory of types to resolve the paradoxes in the foundations of mathematics.

One pillar of type theory is that functions differ from individuals. There are also functions of functions, etc., giving a hierarchy of types. To start, there is a type of individuals (Church's $\iota$) and a type of propositions (Church's $o$). If $\alpha$ and $\beta$ are types then so is $\alpha \to \beta$, the type of functions from $\alpha$ to $\beta$. A statement is meaningless unless it obeys the type constraints: a function of type $\alpha \to \beta$ can only be applied an object of type $\alpha$. The logical constants are functions over the type of propositions. In $\forall x$ and $\exists x$ the variable $x$ must range over some type.

Russell's other pillar was the *vicious circle principle*, which concerned statements like 'all propositions are either true or false' [37, page 37]. If this were itself a proposition then it would refer to itself, a possibly dangerous circularity. The vicious circle principle forbade such 'impredicative propositions' through a system of orders. A statement about all propositions of order $n$ was itself a proposition of order $n+1$. Whitehead and Russell showed how this ramified type theory resolved the paradoxes, but it was too weak to justify classical mathematics. They were forced to assume the Axiom of Reducibility, squashing the orders down to one.

Simple type theory remains when the vicious circle principle is abandoned. Although sets must be introduced in a strict hierarchy, propositions need not be. The idea of orders appears today in the *universes* of Martin-Löf's Type Theory [22], where propositions are represented by types. The terminology persists: simple type theory is called *higher-order logic* because it permits unrestricted quantification over propositions of all orders.

The main achievement of Church [5] is a precise formulation of the syntax. (Gödel calls the vague syntax in *Principia* 'a considerable step backwards as compared with Frege' [13, page 448].) Church formalizes syntax, including quantifiers, in the typed $\lambda$-calculus. His technique is now standard in generic theorem proving.

See Hatcher [16] and Gödel [13] for further discussion of the history of these type theories, and Andrews [1] for the formal development.

## 4  Fundamental issues in type theory

The following sections discuss basic issues in type theory: subtypes, description operators, empty types, polymorphism, and higher-order reasoning. Here, we consider the semantics and syntax.

We begin with basic notation and conventions for syntactic variables:

- *types* are Greek letters $\alpha$, $\beta$, $\gamma$, ...

- *bound variables* are $x$, $y$, $z$

- *terms* are $a$, $b$, $c$, $d$, ...; the ordered pair of $a$ and $b$ is $\langle a, b \rangle$; the relation $a : \alpha$ means '$a$ has type $\alpha$'

- *formulae* are $P$, $Q$, $R$, ...; the true formula is $\top$; the false formula is $\bot$

### 4.1  Types as sets

Type theory is intended as the foundation of mathematics, but it has a simple interpretation in set theory. Types denote sets, abstractions denote set-theoretic functions, and the typing relation denotes set membership. The present formulation retains this semantics. There is no clear alternative: most mathematical reasoning involves sets.

Many theorem provers perform type checking with parsing, but Isabelle cannot do this for its object-logics. Type errors are detected much later: during proofs. Proofs must include explicit type checking using type inference rules, like in Martin-Löf's Type Theory. Type symbols appear as extra arguments to constants.

Ill-typed terms can be written. Their value can vary among models because it is not determined by the axioms. There is no special 'undefined' value. Similarly, an ill-typed formula has some truth value. If this seems unsatisfactory, observe that a traditional theory of Peano arithmetic specifies no value for division by zero, yet the term $a/0$ denotes some number in each model, for each $a$.

Of course, there are alternative semantics. Fourman and Scott [11, 33] can reason about whether $a/b$ exists, but their *existence predicate* involves some complexity. Their logic has a topos semantics, which is a categorical generalization of set theory. Martin-Löf's Type Theory [22] has a constructive, operational semantics. An ambitious type theory can even be based on classical sets: Borzyszkowski et al. formalize general products, some domain theory, and types of types [10]. If the present logic seems pedestrian compared with these, remember that it claims to express the Simple Theory of Types.

## 4.2  Variable binding and substitution

Isabelle has a typed $\lambda$-calculus at the meta-level to deal with operators, variable binding, and substitution uniformly.[1] All forms of variable binding — abstraction, quantifiers, descriptions — are expressed through meta-level abstraction. Compound expressions like $\text{fst}(a)$ and $P \mathbin{\&} Q$ are expressed through meta-level application. Further details are discussed elsewhere [30]. It may be simpler to regard all compound forms, variable-binding or not, as primitive. When a term is written as $b(x)$, this can be regarded as setting off the occurrences of $x$, so that $b(a)$ nearby indicates substitution. In fact, $b(x)$ is a meta-level application, and substitution takes place by meta-level $\beta$-reduction.

Church represents syntax in the object-level typed $\lambda$-calculus. His formulation defines $\lambda$-abstraction and application, then uses these to express quantifiers and descriptions.

Meta-abstraction works better than object-abstraction in Isabelle. It is also more modular. With meta-abstraction always available, different fragments of the logic

---
[1]This works like Martin-Löf's system of arities [26].

can be understood independently of its internal notion of function.

A meta-level function may not correspond to any object-level function. For example, the pairing function is defined for all values of all types. It is defined over the whole object-level domain. It cannot be an object-level function, but could be represented by a family of object-level functions of various types. This must be borne in mind when comparing the present formulation with Church's.

**Notation for object-level functions.** Abstraction is written $\lambda x : \alpha \, . \, b(x)$. Application is written with the explicit 'apply' operator ('), as in $f \, ` \, a$. The apply operator is not used in general discussions of simple type theory.

## 5 Subtypes

A subtype is a collection of the elements of a type that share some common property. Typically, a subtype defines an abstract type from a type of representations. The abstract type contains just the elements that represent abstract objects. Any predicate $P(x)$ over a type $\alpha$ defines a subtype $\{x : \alpha \, . \, P(x)\}$. Subtypes usually make type checking undecidable, for checking whether $a$ belongs to $\{x : \alpha \, . \, P(x)\}$ requires proving $P(a)$.

Consider defining the sum type $\alpha + \beta$. The left injection of $a$ can be represented by the pair of abstractions

$$\langle (\lambda x : \alpha \, . \, a = x), (\lambda y : \beta \, . \, \bot) \rangle$$

while the right injection of $b$ can be represented by

$$\langle (\lambda x : \alpha \, . \, \bot), (\lambda y : \beta \, . \, b = y) \rangle$$

If formulae are terms of type *bool*, both injections have type $(\alpha \to bool) \times (\beta \to bool)$. The subtype containing just the injections is the sum type $\alpha + \beta$.

Because a subtype may depend upon bound variables, we must consider introducing dependent types: general products and sums. These cause no semantic difficulties, but seem unnecessary (see also Dana Scott [33]). The term

$$\lambda z : \alpha \, . \, \lambda y : \{x : \alpha \, . \, R(z, x)\} \, . \, y$$

has no legal type. Its type could be $\prod_{z:\alpha} . \{x : \alpha . R(z,x)\} \to \alpha$ if we added dependent types. However

$$\lambda z : \alpha \, . \, \exists y : \{x : \alpha \, . \, R(z, x)\} \, . \, P(y)$$

has type $\alpha \to \textit{bool}$ because the body of the abstraction is a formula.

Traditionally a term has a unique type, but each element of a subtype also belongs to its parent type. Gödel [13, page 466] describes uniqueness of types as a suspect principle, noting that it precludes reasoning about types. Such reasoning is a necessity for Isabelle. Allowing types to overlap causes no difficulty in the set-theoretic semantics.

Gordon's HOL has a different treatment of subtypes (see Melham [23]). To keep uniqueness of types and decidable type checking, conversion functions distinguish elements of a subtype from elements of the parent type. Determining that the conversion functions are applied correctly still requires theorem proving. Subtypes are defined by top-level commands, so there is no question of dependent types. Subtypes must be non-empty.

# 6 Descriptions

Descriptions, present in type theory from the beginning, name an object by a defining property. The unique description $\iota x : \alpha \, . \, P(x)$ means 'the $x$ satisfying $P(x)$'. For example, $\sqrt{a}$ is $\iota x : \textit{nat} \, . \, x^2 = a$. The inference rule verifies that there exists a unique value that satisfies $P(x)$:

$$\frac{\exists x : \alpha \, . \, P(x) \, \& \, (\forall y : \alpha \, . \, P(y) \to y = x)}{P(\iota x : \alpha \, . \, P(x))}$$

Descriptions can also embody the Axiom of Choice. Hilbert's $\epsilon$-operator, written $\epsilon x : \alpha \, . \, P(x)$, means 'some $x$ satisfying $P(x)$'. The rule drops the requirement of uniqueness.

$$\frac{\exists x : \alpha \, . \, P(x)}{P(\epsilon x : \alpha \, . \, P(x))}$$

Replacing the premise $\exists x : \alpha \, . \, P(x)$ by the two premises $a : \alpha$ and $P(a)$ would impose the stronger requirement (especially in classical logic) of exhibiting the term $a$.

Gordon's HOL uses Hilbert's $\epsilon$-operator, while Church [5] formalizes both forms of description. See Leisenring [21] for a full discussion of Hilbert's $\epsilon$-operator.

## 6.1 Descriptions in *Principia Mathematica*

Whitehead and Russell (in Chapter III of *Principia*) argue that descriptions are meaningless by themselves. They give translations to eliminate descriptions from statements. In their view *The author of Waverley was a poet* means Waverley was written by some poet, not that Sir Walter Scott was a poet. This is because if *The author of Waverley* denotes Sir Walter Scott, then *Scott is the author of Waverley* means the same as *Scott is Scott*, which cannot be intended. Girard calls this the question of *sense vs denotation* [12]. The question applies broadly, not just to descriptions: does $2+2=4$ mean simply $4=4$?

If there is no object meeting the description, its meaning is problematical. To Whitehead and Russell, a statement like *The present King of France is bald* is simply false, while the meaning of *The present King of France is not bald* depends upon the scope of the *not*.

The modern view is that a description denotes some object satisfying the given property, if there is one. Otherwise it is undefined — however we understand this.

## 6.2 The Axiom of Choice and classical logic

In higher-order logic, the Axiom of Choice implies the excluded middle. The argument, due to Diaconescu, is sketched by D. Scott [33]. Let *two* be the type whose values are 0 and 1. To derive $P \vee \neg P$ for some formula $P$, define type *set* as the following subtype of $two \to bool$:

$$set \equiv \{q : two \to bool \,.\, \exists x : two \,.\, q(x) \,\&\, ((\forall x : two \,.\, q(x)) \leftrightarrow P)\}$$

Note that $q_0$ and $q_1$ belong to type *set*, where

$$q_0 \equiv \lambda x : two \,.\, (x = 0) \vee P$$
$$q_1 \equiv \lambda x : two \,.\, (x = 1) \vee P$$

Informally, $q_0$ and $q_1$ correspond to the sets $\{0, 1?\}$ and $\{1, 0?\}$, where 0? and 1? are included just if $P$ holds. Each element of *set* contains some $x : two$. By the Axiom of Choice (Hilbert's $\epsilon$-operator) there is a corresponding function $f : set \to two$:

$$f \equiv \lambda q : set \,.\, \epsilon x : two \,.\, q(x)$$

Whether $f(q_0) = f(q_1)$ holds or not is decidable, for it is an equality between natural numbers. And this equality decides $P$:

- If $f(q_0) = f(q_1)$ then $P$. By the rule for descriptions, both $q_0(f(q_0))$ and $q_1(f(q_1))$ hold, namely $(f(q_0) = 0) \vee P$ and $(f(q_1) = 1) \vee P$. There are four subcases, of which three imply $P$ and one implies $0 = 1$.

- If $f(q_0) \neq f(q_1)$ then $\neg P$. Assuming that $P$ holds implies $q_0 = \lambda x : two \,.\, \top = q_1$, and so $f(q_0) = f(q_1)$, contradiction.

# 7 Empty types

Traditional formulations of higher-order logic require that all types are non-empty. This hardly matters for Church, who has only two basic types (propositions and individuals). With many basic types, the requirement becomes unnatural. If the subtype $\{x : \alpha \,.\, P(x)\}$ depends on free variables, it could sometimes be empty. Empty types require careful formulation of quantifiers and descriptions.

Isabelle's meta-logic uses implicit type checking and does not admit empty meta-types. Empty types are not needed at the meta-level.

## 7.1 Church's formulation of quantifiers

Church postulates a supply of variables $x_\alpha$, $y_\alpha$, $z_\alpha$, ... for each type $\alpha$. Typical quantifier rules are (with the usual variable restrictions)

$$\frac{P(x_\alpha)}{\forall x_\alpha \,.\, P(x_\alpha)} \;\forall\text{-intr} \qquad \frac{\forall x_\alpha \,.\, P(x_\alpha)}{P(a_\alpha)} \;\forall\text{-elim}$$

$$\frac{P(a_\alpha)}{\exists x_\alpha \,.\, P(x_\alpha)} \;\exists\text{-intr} \qquad \frac{\exists x_\alpha \,.\, P(x_\alpha) \quad \begin{array}{c}[P(x_\alpha)]\\ Q\end{array}}{Q} \;\exists\text{-elim}$$

In the $\forall$-intr rule $x_\alpha$ is a variable of type $\alpha$. In the $\forall$-elim rule, $a_\alpha$ is any term of type $\alpha$, even a variable. This rule and $\exists$-intr are unsound if $\alpha$ is empty, with many false consequences:

$$\neg(\forall x_\alpha \,.\, \bot) \qquad \exists x_\alpha \,.\, \top \qquad (\forall x_\alpha \,.\, P(x_\alpha)) \to (\exists x_\alpha \,.\, P(x_\alpha))$$

These look like trivial theorems of first-order logic, but a first-order domain may not be empty.

## 7.2 Quantifier rules admitting empty types

Explicit type checking admits empty types — almost by accident. The quantifier rules are

$$\frac{\begin{array}{c}[x : \alpha]\\ P(x)\end{array}}{\forall x : \alpha \,.\, P(x)} \;\forall\text{-intr} \qquad \frac{\forall x : \alpha \,.\, P(x) \quad a : \alpha}{P(a)} \;\forall\text{-elim}$$

$$\frac{P(a) \quad a : \alpha}{\exists x : \alpha \,.\, P(x)} \;\exists\text{-intr} \qquad \frac{\exists x : \alpha \,.\, P(x) \quad \begin{array}{c}[x : \alpha,\; P(x)]\\ Q\end{array}}{Q} \;\exists\text{-elim}$$

There is no supply of typed variables; instead, ∀-intr and ∃-elim discharge the assumption $x : \alpha$. The rules ∀-elim and ∃-intr demand a proof of $a : \alpha$; if $a$ is some variable $y$, the proof will depend on the assumption $y : \alpha$. If there is no closed term of type $\alpha$ then $\exists x : \alpha \mathbin{.} \top$ has no proof, but $\forall y : \alpha \mathbin{.} \exists x : \alpha \mathbin{.} \top$ does:

$$\frac{\dfrac{\top \quad [y : \alpha]}{\exists x : \alpha \mathbin{.} \top}}{\forall y : \alpha \mathbin{.} \exists x : \alpha \mathbin{.} \top}$$

## 7.3 Descriptions and empty types

Because it is always defined, Hilbert's $\epsilon$-operator gives every type $\alpha$ the element $\epsilon x : \alpha \mathbin{.} \top$. Another form of description, the $\eta$-operator, permits empty types. If there is no $x : \alpha$ satisfying $P(x)$ then $\eta x : \alpha \mathbin{.} P(x)$ is *undefined* — its value and type are unspecified. This typing rule for descriptions makes type checking undecidable.

Hilbert's $\epsilon$-operator can express the quantifiers: for example, $\exists x : \alpha \mathbin{.} P(x)$ as $P(\epsilon x : \alpha \mathbin{.} P(x))$. This does not work with the $\eta$-operator, for if $\exists x : \alpha \mathbin{.} P(x)$ is false then $P(\eta x : \alpha \mathbin{.} P(x))$ is meaningless.

The $\epsilon$-operator can be defined through the $\eta$-operator if $\alpha$ is non-empty, because $Q \to \exists x : \alpha \mathbin{.} P(x)$ implies $\exists x : \alpha \mathbin{.} Q \to P(x)$ under classical logic.[2] Putting $\exists y : \alpha \mathbin{.} P(y)$ for $Q$, the body of the $\eta$ can always be satisfied:

$$\epsilon x : \alpha \mathbin{.} P(x) \;\equiv\; \eta x : \alpha \mathbin{.} (\exists y : \alpha \mathbin{.} P(y)) \to P(x)$$

## 7.4 Alternative formulations

The above quantifier and description rules are adopted for the present formulation of simple type theory. Here are two other ways — both based on topos theory — of admitting empty types.

- Fourman [11] and Dana Scott [33] formalize the notion of existence: a term can have a valid type and yet be undefined. A type is empty if it has no defined elements. The ∀-elimination rule can only be applied to a defined term. The description $\iota x : \alpha \mathbin{.} P(x)$ exists only if $P(x)$ is satisfied by a unique value, but always has type $\alpha$.

- Lambek and P. J. Scott [20, page 130] present quantifier rules that maintain a list of the typed variables on which the conclusion depends.

---

[2]Classical logic obtains by Diaconescu's argument, for $\eta$ implies the Axiom of Choice.

# 8 Polymorphism

The 'typical ambiguity' in *Principia* is a form of polymorphism where type symbols in expressions are simply not shown. Type checking in ML is a formal version of the same thing: types are inferred but not shown. Isabelle does not (at present) allow any hiding of syntax. This calls for another kind of polymorphism, where certain constants have no type symbols at all. Let us consider how to minimize type symbols in elements of function, product, and sum types. To do this safely, we must remember the semantics.

## 8.1 Functions

The notation for abstraction could have type symbols for the function's domain and range:
$$\lambda_{\alpha,\beta} x \, . \, b(x) \quad : \quad \alpha \to \beta$$
where $b(x) : \beta$ for $x : \alpha$.

The type symbol $\alpha$ is essential. Functions like $\lambda x \, . \, 0$ and $\lambda x \, . \, x$ cannot be interpreted as sets without specifying the set of values $x$ may take. In domain theory a function can be defined over the universal domain. But we are in set theory, where an operation 'on everything' is not a function. An element of $\alpha \to \beta$ may (hereditarily) contain all the elements of $\alpha$ and $\beta$, so these must be sets.

The type symbol $\beta$ is superfluous, however, by set theory's Axiom of Replacement. If $\alpha$ denotes a set then $\{b(x) \mid x : \alpha\}$ is also a set. Deleting $\beta$ improves the notation:
$$\lambda x : \alpha \, . \, b(x) \quad : \quad \alpha \to \beta$$

Martin-Löf's Type Theory has polymorphic functions like $\lambda x . x$, but its semantics is operational: its functions are algorithms, not graphs. Nor can we omit the type symbols as a syntactic convenience, hoping they could be replaced in principle. Anne Salvesen [32] presents proofs in Martin-Löf's Type Theory that fail when type symbols are added. Her arguments are general, and should apply here as well.

Note that only the 'contravariant' aspect of functions — the domain of application — requires a type label. Product and sum types do not need any type labels.

## 8.2 Products

For products, a pairing constructor with type symbols is
$$\text{Pair}_{\alpha,\beta}(a)(b) : \alpha \times \beta \qquad (a : \alpha, b : \beta)$$

This could abbreviate the abstraction

$$\lambda x : \alpha \,.\, \lambda y : \beta \,.\, (x = a) \,\&\, (y = b) \quad : \quad \alpha \to \beta \to bool$$

which contains the type symbols $\alpha$ and $\beta$. But pairs need not carry type symbols. In set theory, pairs formed by the operation $\langle a, b \rangle = \{\{a\}, \{a, b\}\}$ do not depend on what sets contain $a$ and $b$. The Isabelle formulation includes a polymorphic pairing operator $\langle a, b \rangle$.

Many other authors, for various reasons, take pairing as a primitive of the typed $\lambda$-calculus [12, 20, 33].

## 8.3 Sums

The disjoint sum has left and right injections:

$$\text{Inl}_{\alpha,\beta}(a) : \alpha + \beta \qquad (a : \alpha)$$

$$\text{Inr}_{\alpha,\beta}(b) : \alpha + \beta \qquad (b : \beta)$$

The type symbols are again unnecessary. In set theory, the injections $\text{Inl}(a) = \langle \{a\}, \emptyset \rangle$ and $\text{Inr}(b) = \langle \emptyset, \{b\} \rangle$ depend only on the values of $a$ and $b$, not on the sets that contain them.

In the current version of the logic, a monomorphic (type labelled) disjoint sum is derived as shown in Section 5. Taking polymorphic injections as primitive seems to be needless extra complexity, for it does not greatly improve the notation.

## 8.4 Comparison with other type systems

The Edinburgh Logical Framework, a type theory for representing formal systems, can express the implicit type checking of Church's higher-order logic [15]. In this case all constants are fully decorated with type symbols, and terms contain much redundant type information.

Gordon's HOL system, like LCF, uses polymorphic type checking. Its type variables, written *, **, etc., are syntactic variables ranging over types. The identity combinator I might have the polymorphic type

```
I : * -> *
```

Because a constant's type is part of its name, the HOL constant I stands for a family of constants $I_\alpha$, satisfying the schematic typing

$$I_\alpha : \alpha \to \alpha$$

For example, I(I) abbreviates $I_{(\alpha\to\alpha)\to(\alpha\to\alpha)}(I_{\alpha\to\alpha})$. This is not self-application: it involves two different instances of $I$.

Under the set-theoretic semantics, an identity function on all types could not exist. Coquand [8] has shown that polymorphic higher-order logic is inconsistent. A constant like $I_\alpha : \alpha \to \alpha$ is here called *monomorphic* (following Salvesen [32]) because of its type label.

# 9 Higher-order reasoning

Quantification over propositions is what makes simple type theory 'higher-order'. First-order logic allows quantification over individuals; second-order logic allows quantification over properties of individuals; third-order logic allows quantification over properties of properties of individuals; and so forth. Higher-order (or $\omega$-order) logic allows all these quantifications. A formula is simply a term of type *bool*.

The logic programming language $\lambda$Prolog, though based on higher-order logic, forbids quantification over *bool* [24]. So it is really first-order logic extended with typed $\lambda$-expressions. The meta-logic of Isabelle avoids quantification over *bool* to simplify the theory, but this restriction is not enforced.

Quantification over propositions permits many different formulations of higher-order logic. Absurdity ($\bot$) is definable as $\forall p : bool \,.\, p$, the proposition that implies all propositions. Conjunction and disjunction are definable by

$$P \,\&\, Q \;\equiv\; \forall r : bool \,.\, (P \to (Q \to r)) \to r$$
$$P \lor Q \;\equiv\; \forall r : bool \,.\, (P \to r) \to ((Q \to r) \to r)$$

Andrews [1] presents a formulation based on equality. For example, the universal quantifier is defined in terms of truth ($\top$) as

$$\forall x : \alpha \,.\, P(x) \;\equiv\; (\lambda x : \alpha \,.\, P(x)) = (\lambda x : \alpha \,.\, \top)$$

Representing formulae by terms of type *bool* is inconvenient in Isabelle. Explicit type inference (to ensure that all theorems have type *bool*) encumbers proofs. Type checking can be minimized by formulating each rule such that the conclusion is well-typed provided its premises are. Then type checking only takes place when assumptions are discharged. In a trial implementation, even this much checking was inefficient.

Now formulae are a separate syntactic class. The present formulation defines first-order logic. It then adds *reflection* — isomorphisms between formulae and terms of type *bool* — to obtain higher-order logic.

- $term(P)$ is a term (of type *bool*) if $P$ is a formula

- $form(b)$ is a formula if $b$ is a term

A predicate (or *class*) on $\alpha$ is just a function of type $\alpha \to bool$. Class formation is $\lambda$-abstraction over a formula, while the membership predicate is function application. Let us introduce some class-theoretic notation:

$$\begin{aligned} \{\!|x:\alpha\,.\,P(x)|\!\} &\equiv \lambda x:\alpha\,.\,term(P(x)) \\ a \in S &\equiv form(S\,`\,a) \end{aligned}$$

Class theory is the main vehicle for mathematical reasoning in *Principia*.

The higher-order logic of Fourman and D. Scott [11, 33] also distinguishes between terms and formulae. The primitive types are products and powersets; functions are represented by their graphs, like in set theory. Reflection functions can be defined through class abstraction and the membership predicate: abstraction creates a class (a term) from a formula, while membership in a class is a formula.

## 10 Recursive data types

Recursive types, like the natural numbers, lists, and trees, are an active research area. The wellordering types of Martin-Löf's Type Theory are general transfinite trees [22]. The Nuprl system, although largely based on Martin-Löf, uses positive recursive type definitions [7]. Boyer and Moore's 'shell principle' introduces recursive structures [4]. LCF can define recursive types using domain theory [29]. Recursive types can also be constructed in simple type theory.

The natural numbers can be constructed in various ways, assuming an Axiom of Infinity. In *Principia*, the number 2 is the class of all pairs of some type $\alpha$. In Church, 2 is $\lambda f:\alpha \to \alpha\,.\,\lambda x:\alpha\,.\,f(fx)$. Both definitions are cumbersome and entail different types of natural numbers for each type $\alpha$. The Isabelle formulation postulates a type *nat* of natural numbers satisfying induction and primitive recursion.

Melham [23] describes one treatment of recursive types, defining lists in terms of natural numbers, and trees in terms of lists. He has implemented this in Gordon's HOL system, which uses Church's logic.

The Isabelle formulation uses a different treatment inspired by Huet [19]. For a given set of constructors it involves two steps:

1. Find a type rich enough to represent all possible constructions.

2. Restrict to the subtype inductively generated by the constructors.

Let us define $list(\alpha)$, the type of lists over $\alpha$. The representing type is $nat \times \alpha \to bool$, where the list $[x_0, x_1, \ldots, x_n]$ is represented by the class of pairs

$$\{\langle 0, x_0 \rangle, \langle 1, x_1 \rangle, \ldots, \langle n, x_n \rangle\}$$

The constructors are $\text{nil}_\alpha$ and $\text{cons}_\alpha(a, l)$. (They must be labelled with type $\alpha$ because classes have type labels.) The empty list is represented by the empty class. To put an element in front of a list, $\text{cons}_\alpha(a, l)$ increments $m$ in all pairs $\langle m, x \rangle$ in $l$, then adds the pair $\langle 0, a \rangle$.

$$\begin{aligned}
\text{nil}_\alpha &\equiv \{u : nat \times \alpha \,.\, \bot\} \\
\text{cons}_\alpha(a, l) &\equiv \{u : nat \times \alpha \,.\, u = \langle 0, a \rangle \\
&\qquad \vee\, (\exists m : nat \,.\, \exists x : \alpha \,.\, \langle m, x \rangle \in l \,\&\, u = \langle \text{Succ}(m), x \rangle)\}
\end{aligned}$$

The representing type includes many non-lists. Tarski's theorem (see Huet [19]), which asserts that every monotone function over a complete lattice has a least fixed point, can be used to define the subtype of lists. The monotone function takes a class $F$ of lists and returns the class of all lists obtained by a further application of the constructors:

$$\begin{aligned}
\{l : nat \times \alpha \to bool \,.\, l = \text{nil}_\alpha \\
\vee\, (\exists x : \alpha \,.\, \exists l' : nat \times \alpha \to bool \,.\, l' \in F \,\&\, l = \text{cons}_\alpha(x, l'))\}
\end{aligned}$$

Trees with labelled edges can also be represented by classes of pairs. For lists, the number in each pair gives the position of an element. The position of an element in a tree can be given by a list of edges. If the trees have countable branching, the representing type could be $list(nat) \times \alpha$. Trees are sometimes represented like this in set theory.

Tarski's theorem also handles recursively defined classes. For example, the reflexive/transitive closure of the relation $R$ is inductively generated from the identity relation by composition with $R$. This too is the least fixed point of a monotone function.

# 11 The formulation of simple type theory

Because of type inference there are two forms of judgement: 'formula $P$ is true', written simply $P$, and 'term $a$ has type $\alpha$', written $a : \alpha$. Type assertions cannot be combined by logical connectives — which would not be in the spirit of type theory — because $a : \alpha$ is not a formula.

Appendix A is the Isabelle rule file, including a few uninteresting rules omitted below.

## 11.1 Equality

The formula $a =_\alpha b$ means that $a$ and $b$ are equal and have type $\alpha$. There is no deep reason for having a typed equality relation. A proof that $a$ equals $b$ must involve showing that $a$ and $b$ have some type $\alpha$, and this type information could be useful later.[3]

The reflexivity, symmetry, and substitution rules are

$$\frac{a : \alpha}{a =_\alpha a} \qquad \frac{a =_\alpha b}{b =_\alpha a} \qquad \frac{a =_\alpha b \quad P(b)}{P(a)}$$

The type information can be extracted. If $a = b$ then both terms have the relevant type.

$$\frac{a =_\alpha b}{a : \alpha} \qquad \frac{a =_\alpha b}{b : \alpha}$$

## 11.2 Types

**Functions**

These rules for abstraction and application are typical of type inference systems: see Chapter 15 of Hindley and Seldin [17]. Applications are written with an explicit operator: $f \, ` \, a$.

$$\frac{\begin{array}{c}[x : \alpha]\\ b(x) : \beta\end{array}}{(\lambda x : \alpha \, . \, b(x)) : \alpha \to \beta} \qquad \frac{f : \alpha \to \beta \quad a : \alpha}{f \, ` \, a : \beta}$$

We have $\beta$ and $\eta$-conversion:

$$\frac{\begin{array}{cc} & [x : \alpha] \\ a : \alpha & b(x) : \beta\end{array}}{(\lambda x : \alpha \, . \, b(x)) \, ` \, a =_\beta b(a)} \qquad \frac{f : \alpha \to \beta}{\lambda x : \alpha \, . \, f \, ` \, x =_{\alpha \to \beta} f}$$

In $\eta$-conversion, variable $x$ may not be free in $f$. All rules that discharge the assumption $x : \alpha$ are subject to the proviso that $x$ is not free in the conclusion or other assumptions. This will be taken for granted below.

Finally, there is a rule for the construction of equal abstractions. It does not follow from the substitution rule above because $x$ is bound in the conclusion.

$$\frac{\begin{array}{c}[x : \alpha]\\ b(x) =_\beta c(x)\end{array}}{(\lambda x : \alpha \, . \, b(x)) =_{\alpha \to \beta} (\lambda x : \alpha \, . \, c(x))}$$

---

[3] In Martin-Löf's Type Theory, equality can only be understood with respect to some type, so the relation is typed for semantic reasons.

### Products

The pair of $a$ and $b$ is written $\langle a, b \rangle$; the projections are fst and snd. These constants contain no type symbols.

Type assignment rules for pairing and the projections are

$$\frac{a : \alpha \quad b : \beta}{\langle a, b \rangle : \alpha \times \beta} \qquad \frac{p : \alpha \times \beta}{\text{fst}(p) : \alpha} \qquad \frac{p : \alpha \times \beta}{\text{snd}(p) : \beta}$$

Conversion (equality) rules for pairing and the projections are

$$\frac{a : \alpha \quad b : \beta}{\text{fst}(\langle a, b \rangle) =_\alpha a} \qquad \frac{a : \alpha \quad b : \beta}{\text{snd}(\langle a, b \rangle) =_\beta b}$$

The elimination rule for products resembles a rule of Martin-Löf's Type Theory:

$$\frac{p : \alpha \times \beta \quad \begin{array}{c} [x : \alpha,\ y : \beta] \\ Q(\langle x, y \rangle) \end{array}}{Q(p)}$$

It implies $\langle \text{fst}(p), \text{snd}(p) \rangle =_{\alpha \times \beta} p$ for $p : \alpha \times \beta$.

### Subtypes

The type checking of subtypes involves the truth of $P(a)$ and is therefore undecidable.

$$\frac{a : \alpha \quad P(a)}{a : \{x : \alpha\ .\ P(x)\}}$$

The elimination rules say that if $a : \{x : \alpha\ .\ P(x)\}$ then $a : \alpha$ and $P(a)$.

$$\frac{a : \{x : \alpha\ .\ P(x)\}}{a : \alpha} \qquad \frac{a : \{x : \alpha\ .\ P(x)\}}{P(a)}$$

### Natural numbers

The type of natural numbers is called *nat*. The Axiom of Infinity is expressed in the most convenient form: through the existence of functions defined by primitive recursion.

The typing rules for 0 and successor are

$$0 : nat \qquad \frac{a : nat}{\text{Succ}(a) : nat}$$

The typing rule for rec is

$$\frac{a : nat \quad b : \beta \quad \begin{array}{c} [x : nat,\ y : \beta] \\ c(x, y) : \beta \end{array}}{\text{rec}(a, b, xy\ .\ c(x, y)) : \beta}$$

The operator $\text{rec}(a, b, xy \,.\, c(x,y))$, where $x$ and $y$ are bound in $c(x,y)$, expresses primitive recursion. In the meta-level typed $\lambda$-calculus $c$ is a function, so $\text{rec}(a, b, xy \,.\, c(x,y))$ will henceforth be abbreviated to $\text{rec}(a,b,c)$. The conversion rules for rec are

$$\frac{b : \beta \quad \begin{array}{c}[x : nat,\ y : \beta]\\ c(x,y) : \beta\end{array}}{\text{rec}(0, b, c) =_\beta b} \qquad \frac{a : nat \quad b : \beta \quad \begin{array}{c}[x : nat,\ y : \beta]\\ c(x,y) : \beta\end{array}}{\text{rec}(\text{Succ}(a), b, c) =_\beta c(a, \text{rec}(a,b,c))}$$

Because rec binds variables, it requires its own substitution rule:

$$\frac{a =_{nat} d \quad b =_\beta e \quad \begin{array}{c}[x : nat,\ y : \beta]\\ c(x,y) =_\beta f(x,y)\end{array}}{\text{rec}(a,b,c) =_\beta \text{rec}(d,e,f)}$$

The mathematical induction rule is

$$\frac{a : nat \quad Q(0) \quad \begin{array}{c}[x : nat,\ Q(x)]\\ Q(\text{Succ}(x))\end{array}}{Q(a)}$$

## 11.3 Logic

Implication and universal quantification are taken as primitive; the other logical constants are defined through them.

The rules for implication are

$$\frac{\begin{array}{c}[P]\\ Q\end{array}}{P \to Q} \qquad \frac{P \to Q \quad P}{Q}$$

The rules for universal quantification are

$$\frac{\begin{array}{c}[x : \alpha]\\ P(x)\end{array}}{\forall x : \alpha \,.\, P(x)} \qquad \frac{\forall x : \alpha \,.\, P(x) \quad a : \alpha}{P(a)}$$

The following rule gives classical logic (which follows anyway from the Axiom of Choice).

$$\frac{\begin{array}{c}[\neg P]\\ P\end{array}}{P}$$

**Reflection**

The operator term($P$) maps a formula to a term of type *bool*, while form($a$) maps such a term to a formula. Since there is no way to decide whether a formula is true or false, term($P$) is non-constructive. The truth value of form($a$) is specified only where $a$ has type *bool*.

The typing rule says that term($P$) has type *bool*, even if $P$ is ill-typed!

$$\text{term}(P) : bool$$

Isomorphism rules state that term and form preserve truth:

$$\frac{a : bool}{\text{term}(\text{form}(a)) =_{bool} a} \qquad \frac{P}{\text{form}(\text{term}(P))} \qquad \frac{\text{form}(\text{term}(P))}{P}$$

Although *form*($a$) is syntactically a formula for all terms $a$, it preserves truth only if $a$ has type *bool*.

Also, term and form preserve equivalence:

$$\frac{\begin{array}{cc}[P] & [Q]\\ Q & P\end{array}}{\text{term}(P) =_{bool} \text{term}(Q)}$$

The analogous property for *form* — that $a =_{bool} b$ and form($b$) imply form($a$) — follows by substitution.

**Definitions of other connectives**

These definitions of other connectives yield their usual properties. The terms False and True have type *bool*; the absurdity formula ($\bot$) is form(False).

$$\begin{aligned}
\text{False} &\equiv \text{term}(\forall p : bool\,.\,\text{form}(p))\\
\text{True} &\equiv \text{term}(\forall p : bool\,.\,\text{form}(p) \to \text{form}(p))\\
P\,\&\,Q &\equiv \forall r : bool\,.\,(P \to Q \to \text{form}(r)) \to \text{form}(r)\\
P \vee Q &\equiv \forall r : bool\,.\,(P \to \text{form}(r)) \to (Q \to \text{form}(r)) \to \text{form}(r)\\
\exists x : \alpha\,.\,P(x) &\equiv \forall r : bool\,.\,(\forall x : \alpha\,.\,P(x) \to \text{form}(r)) \to \text{form}(r)\\
\neg P &\equiv (P \to \text{form}(\text{False}))\\
P \leftrightarrow Q &\equiv (P \to Q)\,\&\,(Q \to P)
\end{aligned}$$

**Descriptions**

The $\eta$-operator is adopted, which assumes an Axiom of Choice and is only defined if some suitable object exists.

$$\frac{\exists x:\alpha\ .\ P(x)}{(\eta x:\alpha\ .\ P(x)):\alpha} \qquad \frac{\exists x:\alpha\ .\ P(x)}{P(\eta x:\alpha\ .\ P(x))}$$

Two descriptions are equal if they are defined and the formulae are equivalent. The second premise ensures that the description is defined.

$$\frac{\begin{array}{c}[x:\alpha]\\P(x)\leftrightarrow Q(x)\end{array}\quad \exists x:\alpha\ .\ P(x)}{(\eta x:\alpha\ .\ P(x))=_\alpha (\eta x:\alpha\ .\ Q(x))}$$

## 11.4 Definitions of types

These include the empty type *void*, the singleton type *unit*, and the union type $\alpha + \beta$.

$$void \equiv \{p:bool\ .\ \text{form}(\text{False})\} \qquad unit \equiv \{p:bool\ .\ p=_{bool}\text{True}\}$$

The sum type consists of all left injections and right injections.

$$\alpha + \beta \equiv \{w:(\alpha \to bool) \times (\beta \to bool).\quad (\exists x:\alpha\ .\ w = \text{Inl}(\alpha,\beta,x))\\ \vee\quad (\exists y:\beta\ .\ w = \text{Inr}(\alpha,\beta,y))\}$$

Injections are defined in a standard way as pairs of classes [33].[4]

$$\text{Inl}(\alpha,\beta,a) \equiv \langle \lambda x:\alpha\ .\ \text{term}(a=_\alpha x), \lambda y:\beta\ .\ \text{False}\rangle$$
$$\text{Inr}(\alpha,\beta,b) \equiv \langle \lambda x:\alpha\ .\ \text{False}, \lambda y:\beta\ .\ \text{term}(b=_\beta y)\rangle$$

The operator $\text{when}(\alpha,\beta,\gamma,p,c,d)$ performs case analysis on a sum type, where $c$ and $d$ are meta-level functions.[5]

$$\text{when}(\alpha,\beta,\gamma,p,c,d) \equiv \eta z:\gamma.\quad (\forall x:\alpha\ .\ p=_{\alpha+\beta}\text{Inl}(\alpha,\beta,x) \to z=_\gamma c(x))\\ \&\quad (\forall y:\beta\ .\ p=_{\alpha+\beta}\text{Inr}(\alpha,\beta,y) \to z=_\gamma d(y))$$

These operators have type labels because they are defined by terms containing type symbols. All variables on the right side in a definition must be present on the left.

Basic laws like $\text{when}_{\alpha,\beta,\gamma}(\text{Inl}_{\alpha,\beta}(a),c,d) = c(a)$ are proved in the Isabelle theory. The operator is computable despite being defined by description.

---

[4]The definition of $\alpha + \beta$ used by Melham [23] does not work if either type is empty.

[5]Conventional notation is $\text{when}_{\alpha,\beta,\gamma}(p, x\ .\ c(x), y\ .\ d(y))$, where $x$ and $y$ are bound variables.

## 11.5 Class Theory

Class theory includes the relations *membership* and *subclass* and the operations *union*, *intersection*, and *powerset*. Union and intersection are also defined for a class of classes. The class abstraction $\{\!|x : \alpha \,.\, P(x)|\!\}$ abbreviates $\lambda x : \alpha \,.\, term(P(x))$. Operators defined by class abstraction have the type label $\alpha$ as an extra argument, so none of these are infix operators in Isabelle's concrete syntax.

$$
\begin{aligned}
a \in S &\equiv form(S \text{`} a) \\
S \subseteq_\alpha T &\equiv \forall z : \alpha \,.\, z \in S \to z \in T \\
S \cup_\alpha T &\equiv \{\!|z : \alpha \,.\, z \in S \vee z \in T|\!\} \\
S \cap_\alpha T &\equiv \{\!|z : \alpha \,.\, z \in S \,\&\, z \in T|\!\} \\
\bigcup\nolimits_\alpha F &\equiv \{\!|z : \alpha \,.\, \exists S : \alpha \to bool \,.\, S \in F \,\&\, z \in S)|\!\} \\
\bigcap\nolimits_\alpha F &\equiv \{\!|z : \alpha \,.\, \forall S : \alpha \to bool \,.\, S \in F \to z \in S)|\!\} \\
\mathcal{P}_\alpha(S) &\equiv \{\!|T : \alpha \,.\, T \subseteq_\alpha S|\!\}
\end{aligned}
$$

# 12 Sample proofs in Isabelle

A logic is traditionally illustrated by sample proofs. Theorems proved using Isabelle include basic facts, lemmas used in proof procedures, Tarski's Theorem, and well-founded recursion. Proof procedures exist for first-order logic, rewriting, and class theory.

## 12.1 Simple proof procedures

The rewriting package is based on the one for Martin-Löf's Type Theory, as are the sample proofs in elementary number theory. Using rewriting and induction, arithmetic is developed up to the theorem $a \bmod b + (a/b) \times b = a$.

Reflection works well in higher-order reasoning. Natural deduction rules for the logical constants are easily derived from their higher-order definitions. A standard example of higher-order logic is Cantor's Theorem that every set has more subsets than elements, which can be expressed as follows:

$$\neg\Big(\exists g : \alpha \to (\alpha \to bool) \,.\, \forall f : \alpha \to bool \,.\, \exists j : \alpha \,.\, f = g \text{`} j\Big)$$

(There is no onto function from $\alpha$ to $\alpha \to bool$.) While TPS [2] can prove Cantor's Theorem automatically, Isabelle must be guided towards the proof.

The proof procedures for first-order logic work directly with the natural deduction rules, as sketched in Chapter 2 of my book [29]. Although none of the procedures is complete or fast, they can prove many examples automatically:

$$\bigl(\exists y:\alpha\,.\,\forall x:\alpha\,.\,J(y,x)\leftrightarrow\neg J(x,x)\bigr) \to \neg\bigl(\forall x:\alpha\,.\,\exists y:\alpha\,.\,\forall z:\alpha\,.\,J(z,y)\leftrightarrow\neg J(z,x)\bigr)$$

Similar proof procedures for class theory reason about unions, intersections, subsets, etc. This example is proved automatically:

$$\frac{F:(\alpha\to bool)\to bool \qquad G:(\alpha\to bool)\to bool}{\bigcap_\alpha(F\cup G)=(\bigcap_\alpha F)\cap(\bigcap_\alpha G)}$$

Classes are also used to construct a type of lists and derive structural induction.

## 12.2 Well-founded induction and recursion

Well-founded recursion is a general method of defining total recursive functions, while well-founded induction reasons about functions so defined. These principles, which hold for every well-founded relation, play a central role in the Boyer/Moore logic [4]. They have been derived using Isabelle.

Given a relation $R:\alpha\times\alpha\to bool$, let us write $y\,R\,x$ instead of $\langle y,x\rangle\in R$ and abbreviate '$R$ is well-founded' as $\mathrm{wf}_\alpha(R)$. Classically, $R$ is well-founded if there are no infinite descending chains $\cdots x_3\,R\,x_2\,R\,x_1\,R\,x_0$. The following definition is more convenient:

$$\begin{aligned}\mathrm{wf}_\alpha(R) \;\equiv\; & \bigl(\forall x:\alpha\,.\,(\forall y:\alpha\,.\,y\,R\,x\to y\in S)\to x\in S\bigr)\\ & \to(\forall x:\alpha\,.\,x\in S)\end{aligned}$$

This easily yields well-founded induction:

$$\frac{\mathrm{wf}_\alpha(R) \qquad a:\alpha \qquad \begin{array}{c}[x:\alpha]\\ \forall y:\alpha\,.\,y\,R\,x\to y\in P(x)\end{array}}{P(a)}$$

A recursive function $f$ is well-founded along $R$ if $f(x)$ depends only on $f(y)$ such that $y\,R\,x$. This condition can be stated using subtypes. Type $\alpha_{Rx}$ is the restriction of $\alpha$ to predecessors of $x$ under $R$.

$$\alpha_{Rx}\equiv\{y:\alpha\,.\,y\,R\,x\}$$

The body of the recursive function has the form $H(x,f)$ where $x:\alpha$ is the argument and $f:\alpha_{Rx}\to\beta$ handles recursive calls. Type checking ensures that $f$ is only called below $x$. The resulting recursive function is applied to argument $a$ by $\mathrm{wfrec}_{\alpha,\beta}(R,H,a)$.

$$\frac{\mathrm{wf}_\alpha(R) \qquad R:\alpha\times\alpha\to bool \qquad a:\alpha \qquad \begin{array}{c}[x:\alpha,\;f:\alpha_{Rx}\to\beta]\\ H(x,f):\beta\end{array}}{\mathrm{wfrec}_{\alpha,\beta}(R,H,a):\beta}$$

Under the same premises, wfrec satisfies the recursion equation

$$\text{wfrec}_{\alpha,\beta}(R, H, a) =_\beta H(a, \lambda x : \alpha_{Ra} \,.\, \text{wfrec}_{\alpha,\beta}(R, H, x))$$

Some observations: The variables $x$ and $f$ are subject to the usual 'not free in' conditions. The abstraction $\lambda x : \alpha_{Ra} \,.\, \text{wfrec}_{\alpha,\beta}(R, H, x)$ restricts the function to arguments below $a$. Here $H$ is a meta-level function (wfrec binds variables); if we had dependent types, $H$ could be an object-function of type $\prod_{x:\alpha}(\alpha_{Rx} \to \beta) \to \beta$.

Defined by a description, wfrec takes the union of all graphs of functions that satisfy the recursion equation below some $x : \alpha$. Its typing rule holds because this union forms the graph of a function on $\alpha$. Observe how type checking can involve substantial proof. With the help of a few extra lemmas, the equality rule is then proved.

This work follows Suppes's treatment of transfinite recursion in set theory [36]. Operator wfrec is defined once and for all, and its properties proved, for all well-founded relations in the logic. It is far stronger than my work in Martin-Löf's Type Theory [28], which considers certain ways of constructing well-founded relations and their corresponding recursion operators.

## 13  Conclusions

Programs are typically verified within a special logic of computation. Although several such logics have been successful, they sometimes restrict abstract mathematical reasoning — needed even for computational proofs.

- The *Logic for Computable Functions* (LCF) embeds a typed $\lambda$-calculus, where types denote domains, into first-order logic [29]. LCF is good for reasoning about nonterminating processes, but termination proofs can become a chore (in my opinion [27]). The restriction to domains and continuous functions has serious consequences [34].

- Martin-Löf's Type Theory is based on computation [22, 26]. By the interpretation of propositions-as-types, a type can express a complete program specification. Developments and applications are proceeding rapidly [3]. However, the theory does not admit classical set-theoretic arguments. Unwanted proof objects in types cause complications [32].

- Boyer and Moore use quantifier-free first-order logic with well-founded induction and recursion [4]. Although this combination gives unique simplicity and power, it is hard to do without quantifiers.

Simple type theory may be suitable for reasoning about computation. It offers a rich collection of computable functions, including general recursive and higher-order functions, but it is not restricted to computable functions. Subtypes and classes can express program specifications. The main question is how to recognize when a function is computable.

Some people will wonder whether classical logic is appropriate. Why not use intuitionistic higher-order logic instead? Simply remove the double-negation law and the Axiom of Choice (replacing the $\eta$-operator by $\iota$). Although I have an interest in constructive logic, this suggestion requires a stronger argument. Intuitionism is a deep and evolving subject. There is little agreement about whether intuitionistic higher-order logic, with its impredicative quantification, is constructive.

The Calculus of Constructions, by Coquand and Huet [9], is also intended for reasoning about programs. In use it is very like simple type theory: the Isabelle proof of Tarski's theorem follows Huet's [19]. However, it interprets propositions-as-types and has a clear notion of computation. Experiments with the Calculus and the Isabelle formulation of type theory will make an interesting comparison.

**Acknowledgement.** Isabelle was developed under grant GR/E 0355.7 from the Science and Engineering Research Council. Mike Fourman made many valuable remarks, especially about descriptions. Thanks also to Thomas Melham, Dale Miller, Philippe Noël, and Jan M. Smith for advice.

# A  Appendix: The Isabelle Rule File

```
(*  Title:  HOL/ruleshell
    Author: Lawrence C Paulson, Cambridge University Computer Laboratory
    Copyright   1989  University of Cambridge

Rules of Higher-order Logic (Type Theory)

!!!After updating, rebuild  ".rules.ML"  by calling make-rulenames!!!
*)
signature HOL_RULE =
  sig
  structure Thm : THM
  val sign: Thm.Sign.sg
  val thy: Thm.theory
(*INSERT-RULESIG -- file produced by make-rulenames*)
  end;

functor HOL_RuleFun (structure HOL_Syntax: HOL_SYNTAX and Thm: THM
        sharing HOL_Syntax.Syntax = Thm.Sign.Syntax) : HOL_RULE =
struct
structure Thm = Thm;

val thy = Thm.enrich_theory Thm.pure_thy "HOL"
    (["term","form","type"], HOL_Syntax.const_decs, HOL_Syntax.syn)
[
  (*** Equality ***)

  ("refl",  "[| a: A |] ==> [| [ a = a : A ]  |]"),

  ("sym",  "[| [ a = b : A ] |] ==> [| [ b = a : A ]  |]"),

  (*Equal terms are well typed -- all rules must enforce this! *)
  ("eq_type1", "[| [ a = b : A ] |] ==> [| a: A |]" ),

  ("eq_type2", "[| [ a = b : A ] |] ==> [| b: A |]" ),

  ("subst",
    "[| [ a = c : A ]  |]  ==>  [| P(c) |]  ==>  [| P(a) |]"),

  (*** TYPES ***)

  (** Functions **)

  ("Lambda_type",
    "(!(x)[| x: A |] ==> [| b(x) : B |]) ==>     \
\    [| lam x:A. b(x) : A->B |]" ),

  ("Lambda_congr",
    "(!(x)[| x: A |] ==> [| [ b(x) = c(x) : B ] |]) ==>     \
\    [| [ lam x:A. b(x) = lam x:A. c(x) : A->B ] |]" ),

  ("apply_type",
    "[| f: A->B |] ==> [| a: A |]  ==> [| f`a : B |]" ),

  ("beta_conv",
```

```
            "[| a : A |] ==> (!(x)[| x: A |] ==> [| b(x) : B |]) ==> \
\           [| [ (lam x:A.b(x)) ' a = b(a) : B ] |]"   ),

       ("eta_conv", "[| f: A->B |] ==> [| [ lam x:A. f'x = f : A->B ] |]"   ),

       (** Products **)

       ("pair_type", "[| a: A |] ==>  [| b: B |]  ==>  [| <a,b> : A*B |]"   ),

       ("prod_elim",
        "[| p : A*B |]   ==> \
\         (!(x,y)[| x: A |] ==> [| y: B |] ==> [| Q(<x,y>) |])  ==> \
\         [| Q(p) |]"    ),

       ("pair_inject",
        "[| [ <a,b> = <c,d> : A*B ] |] ==> \
\         ([| [ a = c : A ] |] ==> [| [ b = d : B ] |] ==> [| R |]) ==> \
\         [| R |]"  ),

       (*fst and snd could be defined using descriptions...they are not to avoid
         excessive type labels -- which is the point of defining products here. *)

       ("fst_type",  "[| p: A*B |] ==> [| fst(p) : A |]"   ),
       ("snd_type",  "[| p: A*B |] ==> [| snd(p) : B |]"   ),

       ("fst_conv",  "[| a: A |] ==>  [| b: B |] ==> [| [ fst(<a,b>) = a: A] |]"   ),
       ("snd_conv",  "[| a: A |] ==>  [| b: B |] ==> [| [ snd(<a,b>) = b: B] |]"   ),

       ("split_def", "split(p,f) == f(fst(p), snd(p))"   ),

       (** Subtypes **)

       ("subtype_intr",  "[| a: A |] ==> [| P(a) |] ==> [| a : {x:A.P(x)} |]"   ),

       ("subtype_elim1",   "[| a: {x:A.P(x)} |] ==> [| a:A |]"),
       ("subtype_elim2",   "[| a: {x:A.P(x)} |] ==> [| P(a) |]"),

       (** Natural numbers **)

       ("Zero_type",  "[| 0: nat |]"   ),
       ("Succ_type",  "[| a: nat |] ==> [| Succ(a) : nat |]"   ),

       ("rec_type",
        "[| a : nat |] ==> \
\         [| b : C |]   ==> \
\         (!(x,y)[| x: nat |] ==> [| y: C |] ==> [| c(x,y): C |])  ==> \
\         [| rec(a,b,c) : C |]"   ),

       ("rec_congr",
        "[| [ a = a' : nat ] |] ==> \
\         [| [ b = b' : C ] |]   ==> \
\         (!(x,y)[| x: nat |] ==> [| y: C |] ==> \
\             [| [ c(x,y) = c'(x,y): C ] |]) ==>   \
\         [| [ rec(a,b,c) = rec(a',b',c') : C ] |]"   ),

       ("rec_conv0",
        "[| b: C |] ==> \
\         (!(x,y)[| x: nat |] ==> [| y: C |] ==> [| c(x,y): C |])  ==> \
\         [| [ rec(0,b,c) = b : C ] |]"   ),

       ("rec_conv1",
        "[| a : nat |] ==> \
\         [| b : C |]   ==> \
\         (!(x,y)[| x: nat |] ==> [| y: C |] ==> [| c(x,y): C |]) ==>   \
```

```
\        [| [ rec(Succ(a),b,c) = c(a, rec(a,b,c)) : C ] |]"   ),

    ("nat_induct",
       "[| a: nat |] ==> [| Q(0) |] ==>     \
\        (!(x)[| x: nat |] ==> [| Q(x) |] ==> [| Q(Succ(x)) |]) ==>   \
\        [| Q(a) |]"   ),

    (*** Logic ***)

    (** Implication and quantification *)

    ("classical",  "([| ~P |] ==> [| P |])  ==> [| P |]"),

    ("imp_intr",
       "([| P |] ==> [| Q |])  ==>   [| P-->Q |]"),

    ("mp",
       "[| P-->Q |] ==> [| P |]   ==> [| Q |]"),

    ("all_intr",
       "(!(x)[| x: A |] ==> [| P(x) |])   ==>   [| ALL x:A.P(x) |]"),

    ("spec",
       "[| ALL x:A.P(x) |] ==> [| a : A |]   ==> [| P(a) |]"),

    (** Reflection *)

    ("term_type", "[| term(P) : bool |]"  ),

    ("term_conv", "[| p: bool |] ==> [| [ term(form(p)) = p : bool ] |]"  ),

    ("form_intr", "[| P |] ==> [| form(term(P)) |]"),

    ("form_elim", "[| form(term(P)) |] ==> [| P |]"),

    ("term_congr",
       "([| P |] ==> [| Q |]) ==> ([| Q |] ==> [| P |]) ==>   \
\        [| [ term(P) = term(Q) : bool ] |]"),

    (** Reduction predicate for simplification. *)

    (*does not verify a:A!  Sound because only trans_red uses a Reduce premise*)
    ("refl_red", "Reduce(a,a)"  ),

    ("red_if_equal", "[| [ a = b : A ] |] ==> Reduce(a,b)"),

    ("trans_red", "[| [ a = b : A ] |] ==> Reduce(b,c) ==> [| [ a = c : A ] |]"),

    (** Definitions of other connectives*)

    ("False_def", "False == term(ALL p:bool.form(p))"),
    ("True_def",  "True == term(ALL p:bool.form(p)-->form(p))"),
    ("conj_def",  "P&Q == ALL r:bool. (P-->Q-->form(r)) --> form(r)"),

    ("disj_def",
     "P|Q == ALL r:bool. (P-->form(r)) --> (Q-->form(r)) --> form(r)"),

    ("exists_def",
     "(EXISTS x:A. P(x)) ==  ALL r:bool. (ALL x:A. P(x)-->form(r)) --> form(r)"),

    ("not_def",   "~P == (P-->form(False))"),
    ("iff_def",   "P<->Q == (P-->Q) & (Q-->P)"),
```

```
   (** Conditionals *)

   ("cond_def", "cond(A,p,a,b) == PICK x:A.(form(p)  & [x=a:A]) | \
\                                       (~form(p) & [x=b:A])" ),

   (** Descriptions *)

   ("Pick_type", "[| EXISTS x:A.P(x) |] ==> [| (PICK x:A.P(x)) : A |]"),

   ("Pick_congr",
     "(!(x)[| x: A |] ==> [| P(x) <-> Q(x) |])   ==>    \
\     [| EXISTS x:A.P(x) |]   ==>  [| [ PICK x:A.P(x) = PICK x:A.Q(x) : A ] |]"),

   ("Pick_intr", "[| EXISTS x:A.P(x) |] ==> [| P(PICK x:A.P(x)) |]"),

   (** Definitions of Classes*)
   ("member_def",       "a<:S == form(S`a)"),
   ("subset_def",       "subset(A,S,T) == ALL z:A. z<:S --> z<:T"),
   ("un_def",           "un(A,S,T) == lam z:A. term(z<:S | z<:T)"),

   ("int_def",          "int(A,S,T) == lam z:A. term(z<:S & z<:T)"),
   ("union_def",
       "union(A,F) == lam z:A. term(EXISTS S:A->bool. S<:F & z<:S)"),
   ("inter_def",
       "inter(A,F) == lam z:A. term(ALL S:A->bool. S<:F --> z<:S)"),
   ("pow_def",
       "pow(A,S) == lam T:A. term(subset(A,T,S))"),

   (** Definitions of types*)

   (*the types "void" and "unit"*)
   ("void_def",  "void == {p: bool. form(False)}"),
   ("unit_def",  "unit == {p: bool. [p=True:bool]}"),

   (*unions: the type A+B *)
   ("plus_def",
     "A+B == {w: (A->bool) * (B->bool). \
\              (EXISTS x:A. [w = Inl(A,B,x) : (A->bool) * (B->bool)]) | \
\              (EXISTS y:B. [w = Inr(A,B,y) : (A->bool) * (B->bool)]) }"),

   ("Inl_def",   "Inl(A,B,a) == <lam x:A.term([ a = x : A ]), lam y:B.False>"),
   ("Inr_def",   "Inr(A,B,b) == <lam x:A.False, lam y:B.term([ b = y : B ])>"),
   ("when_def",
     "when(A,B,C,p,c,d) == PICK z:C.  \
\        (ALL x:A. [ p = Inl(A,B,x) : A+B ] --> [ z = c(x) : C ]) & \
\        (ALL y:B. [ p = Inr(A,B,y) : A+B ] --> [ z = d(y) : C ])"  )];
end;
```